\documentclass{kluwer}

\usepackage{epsfig}

\begin{document}
\begin{article}
\begin{opening}
\title{MSST observations of the pulsating sdB star PG\,1605+072}            

\author{S.~J. \surname{O'Toole}\email{otoole@sternwarte.uni-erlangen.de}}
\author{S. \surname{Falter}}
\author{U. \surname{Heber}} 
\institute{Dr Remeis-Sternwarte, Astronomisches Institut der Universit\"at
Erlangen-N\"urnberg, Sternwartstr.\ 7, Bamberg D-96049, Germany}

\author{C.~S. \surname{Jeffery}}
\institute{Armagh Observatory, College Hill, Armagh BT61 9DG, Northern Ireland, UK}

\author{S. \surname{Dreizler}}
\author{S.~L. \surname{Schuh}}
\institute{Insitut f\"ur Astronomie und Astrophysik, Universit\"at
T\"ubingen, Sand 1, D-72076 T\"ubingen, Germany \\
Universit\"atssternwarte, Universit\"at G\"ottingen, Geismarlandstrasse
11, D-37083 G\"ottingen, Germany}

\author{and the MSST+WET
  \surname{Teams}\thanks{see \texttt{http://astro.uni-tuebingen.de/$\sim$schuh/msst/astronomers.html}
  and \texttt{http://wet.iitap.iastate.edu/xcov22/people.html}}}
\institute{}

\runningtitle{The pulsating sdB star PG\,1605+072}
\runningauthor{S.~J. O'Toole et al.}

\begin{abstract} 
We present the first results from the MultiSite Spectroscopic Telescope
(MSST) observations of the sdBV star PG\,1605+072. Pulsating sdB stars
(also known as EC\,14026 stars) offer the chance to gain new insights
into the formation and evolution of extreme Horizontal Branch stars
using the tools of asteroseismology. PG\,1605+072 is an outstanding
object in its class, with the richest frequency spectrum, the longest
periods, and the largest variations.

The MSST campaign took place in May/June 2002 immediately following
the Whole Earth Telescope Xcov22 run, which observed PG\,1605+072 as an
alternate target. We will first give an overview of the project and
its feasibility, after which we will present the massive data set,
made up of 399 hours of photometry and 151 hours of spectroscopy. The
overall aims of the project are to examine light/velocity amplitude
ratios and phase differences, changes in equivalent width/line index,
and $\lambda$-dependence of photometric amplitudes, and to use these
properties for mode identification.

\end{abstract}

\end{opening}
\section{Motivation}

Of all the short-period sdB pulsators currently known, PG\,1605+072
stands out. It is perhaps the most evolved, it has the highest
amplitudes and longest periods, and it is the only apparently single
sdB star that shows significant rotation. Its rich pulsation spectrum
and relative brightness make it an ideal target for multisite
observing campaigns such as the Whole Earth Telescope (WET). After the
discovery observations showed a complex amplitude spectrum with many
modes (Koen et al.\ 1998), a two week multisite photometric campaign was
organised, and more than 55 frequencies were detected, as well as
evidence for amplitude variation (Kilkenny et al.\ 1999). Models by
Kawaler (1999) suggested that part of the complexity of PG\,1605+072's
amplitude spectrum may be due to rapid rotation with an equatorial
velocity of about 130\,km\,s$^{-1}$, thereby causing rotational
splitting of the oscillation modes. When Heber et al.\ (1999) carried
out a spectral analysis, they measured $v\sin i$ to be
39\,km\,s$^{-1}$, which fit nicely into this picture. Both Heber et
al.\ (1999) and Koen et al.\ (1998) found that PG\,1605+072 has
evolved off the extreme Horizontal Branch (EHB), making it an
important link between the EHB and the white dwarf cooling curve. An
analysis by Reed (2001) showed that the amplitude spectrum is too
complex to measure mode stability.

O'Toole et al.\ (2000) were the first to detect velocity variations in
the Balmer lines of PG\,1605+072, measuring amplitudes of
14\,km\,s$^{-1}$ in H$\beta$ using 2\,m telescopes. Woolf et
al.\ (2002) showed the advantage of using 4\,m class telescopes, with
much better velocity accuracy than 2\,m telescopes. They also used
moments of the cross-correlation function to detect line shape
variations. O'Toole et al.\ (2002), using a larger data set than
O'Toole et al.\ (2000), did a detailed analysis of two-site
spectroscopy, and found evidence for amplitude variation and closely
spaced modes. O'Toole et al.\ (2003) examined Balmer line indices and
found an amplitude dependence on Balmer line number, which they used
to derive the amplitudes of the effective temperature and surface gravity
variations. Falter et al.\ (2003) were the first to attempt
simultaneous multicolour photometry and spectroscopy, and found no
phase difference between different filters. Using
spectrophotometry, O'Toole (2003) obtained the same result, and found
the velocity/intensity phase difference to be $\sim$110$^{\circ}$ for
all 8 modes detected (for purely adiabatic oscillations, this phase
difference should be 90$^{\circ}$).

These analyses showed that one and two site spectroscopic campaigns
are not enough to understand the complex nature of PG\,1605+072, and
that even multisite photometry would not do the job. Considering the
potentially huge amount of information that can be obtained from this
star, we decided to undertake a multisite spectroscopic and
photometric campaign in May/June 2002. Here we present the initial
results from this ambitious project.

\section{Observations and Reductions}

The MSST campaign obtained both photometry and spectroscopy. The
photometric part of the campaign was divided into two parts. As part
of the WET \textsc{Xcov22} campaign, PG\,1605+072 was observed as an
alternative target, and $\sim$127 hours of observations were obtained
(Heber et al.\ 2003). All photomultiplier (PMT) data were reduced
using the WET reduction software QED. Most CCD data were reduced using
standard aperture photometry routines in IRAF. Some data remains
unreduced. From the main part of the MSST campaign $\sim$272 hours of
observations were obtained, giving a total of $\sim$399 hours, or
roughly 54\% temporal coverage. This is by far the most data acquired
for any sdB during a single observing campaign. Again PMT data were
reduced using QED, while most CCD data was reduced using TRIPP (Time
Resolved Imaging Photometry Package, see Schuh et al.\ 1999), and some
data were reduced using IRAF. We also have 5 nights of multicolour
photometry, using BUSCA on the Calar Alto 2.2\,m telescope, which has
yet to be reduced.

\begin{figure}
\begin{center}
\epsfig{file=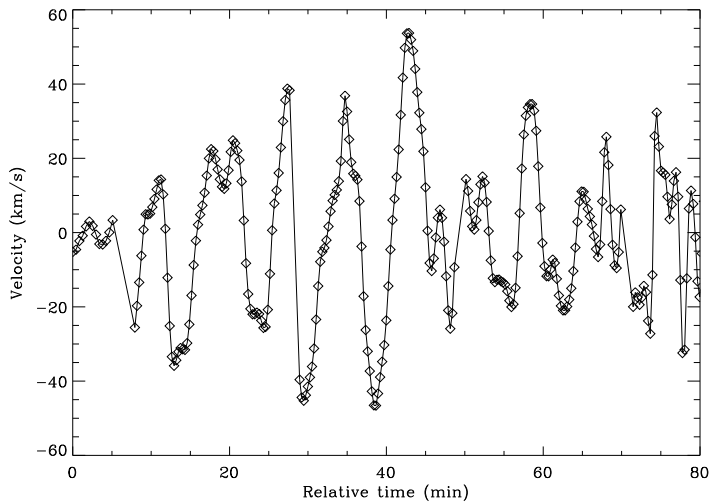,scale=0.56}
\end{center}
\caption{Velocity curve from the Calar Alto 3.5\,m with the TWIN
  spectrograph. Variations with a period of $\sim$8 minutes are
  clear.}
\label{fig:twinvel}
\end{figure}

There were also two parts to the spectroscopic contribution to the MSST
campaign, using 4\,m and 2\,m telescopes. Several of
the 4\,m telescopes applied for did not receive time or were clouded
out, so coverage was poor at only $\sim$7\% (around 27 hours
total). All 3 telescopes (Apache Point 3.5\,m, Calar Alto 3.5\,m and
the ESO NTT 3.5\,m) acquired
spectra by trailing the star along the slit. This allows for variable
exposure times, depending on conditions. An example of the quality of
data possible using this technique is shown in Figure
\ref{fig:twinvel}. The high velocity precision achievable with this
technique suggests that it can be used on fainter
and/or lower amplitude targets. Not all of the data has been analysed,
although everything has been reduced using SPEX (long-slit SPectrum
EXtraction package\footnote{see
\texttt{http://astro.uni-tuebingen.de/$\sim$schuh/spex/index.html}}), a
package which allows for the reduction of trailed spectra. Poor weather
conditions during the NTT observations means that data may not be useful.

\begin{figure}
\begin{center}
\epsfig{file=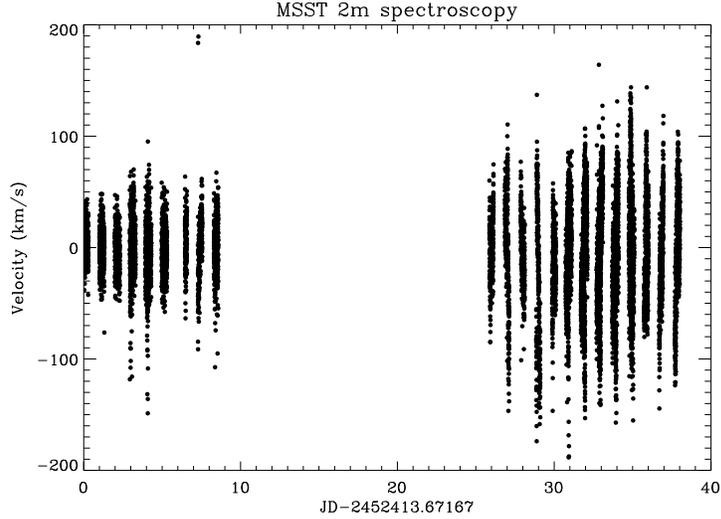,scale=0.56}
\end{center}
\caption{Coverage of the 2\,m spectroscopy part of the campaign. The 2
weeks gap in the observations causes $\sim$0.8\,$\mu$Hz alias peaks as
well as those caused by daily gaps.}
\label{fig:keelecover}
\end{figure}

The observations using 2\,m telescopes were somewhat more successful,
although the timing of the allocations was not always optimum, meaning
there were two halves to the campaign, separated by about 2 weeks. The
first half of the campaign had good coverage on paper, but bad weather
at both La Silla and Siding Spring Observatories meant around 70-75\%
of allocated time was lost, leading to 58 hours of observations or
$\sim$22\% coverage. The second half of the campaign was much more
successful, with 93 hours of observations, or around 32\%
coverage. All of the 2\,m spectroscopy data were reduced using
standard routines in IRAF for bias subtraction, flat fielding, sky
correction, and order extraction, however the velocities were
determined using a double precision version of the rv
package\footnote{available from
\texttt{http://iraf.noao.edu/scripts/extern/rvx.pl}}. The raw
velocities are shown in Figure \ref{fig:keelecover}. The higher
apparent scatter in the second half of the campaign is mainly due to
long term drifts in the velocity curves. These drifts seem to be
inherent in both the DFOSC and ALFOSC spectrographs. A similar problem
is seen in observations of PG\,1325+101 using ALFOSC ({\O}stensen,
these proceedings).

\section{First Results}

\begin{figure}
\begin{center}
\epsfig{file=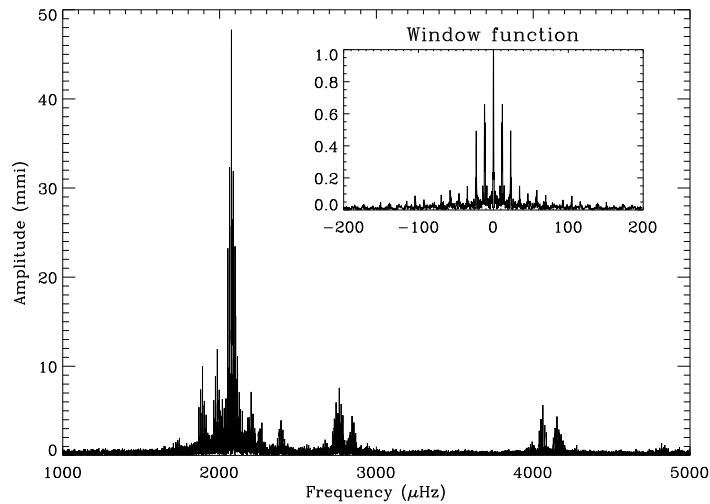,scale=0.56}
\end{center}
\caption{Amplitude spectrum of MSST photometry from SAAO, SSO and the
  JKT. The 2075.8\,$\mu$Hz mode is dominant again. The inset shows the
  spectral window.}
\label{fig:keele3phot}
\end{figure}

Although great care was taken to make sure that the timing of each
observation was accurate, inevitably we ran into some problems. These
mainly occurred during analysis of the photometry, where the addition
of several sites to the main campaign data caused strange aliasing
effects. In some cases a large reduction in oscillation amplitudes was
also seen (up to $\sim$25\%), despite the data from single sites
analysed individually showing similar amplitudes. This might indicate a
timing problem. As a start, we show in Figure \ref{fig:keele3phot} the
amplitude spectrum of 3 sites where timing does not seem to cause
problems (SAAO, JKT and SSO). Fortunately PG\,1605+072 was observed
for at least 6 nights from each of these observatories, constituting a
large fraction of the data. Other sites appear to have deviating
timing, and the reasons are still under investigation. As mentioned
above, some data (from BAO) have not been reduced yet, and are not
included.

One of the most striking things about Figure \ref{fig:keele3phot} is
the dominance of the mode at 2075.8\,$\mu$Hz. This mode had the
highest amplitude in the observations of Koen et al.\ (1998) and
Kilkenny et al.\ (1999), however, in the radial
velocity studies of O'Toole et al.\ (2000, 2002) and Woolf et
al.\ (2002), with observations in 1999 and 2000, it was almost
undetectable or had a much
lower amplitude. It had returned to its former glory by the
time Falter et al.\ (2003) observed it in 2001. Only a quick-and-dirty
analysis of frequencies and amplitudes has been done, and this was
mainly to investigate phase differences between each site. A full
analysis will be done when all data is reduced and the timing problems
are solved. Further discussion of the timing problems and
possible solutions can be found in Section \ref{sec:disc}.
  

An example of what can be achieved with a 1200 lines/mm grating and a
3.5\,m telescope has already been shown in Figure
\ref{fig:twinvel}. Velocity variations are clearly visible with a
period of $\sim$8 minutes. These velocities will be combined with the
2\,m observations once all of the data is fully reduced.

\begin{figure}
\begin{center}
\epsfig{file=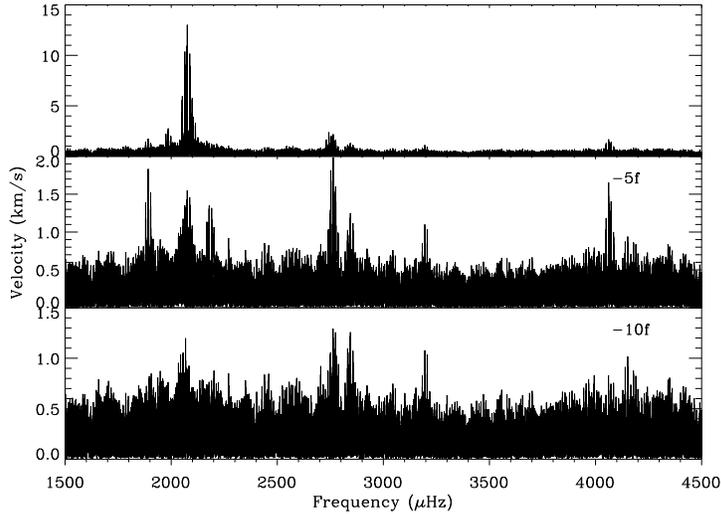,scale=0.56}
\end{center}
\caption{Velocity amplitude spectrum of PG\,1605+072 (\emph{top});
  after prewhitening 5 frequencies (\emph{middle}); after
  prewhiteneing by 10 frequencies (\emph{bottom}).}
\label{fig:keelepw}
\end{figure}

The 2\,m spectroscopic observations appear to be free from timing
difficulties, probably since the number of observatories used was less
than for the photometry. The amplitude spectrum of \textit{all} of the
observations is shown in the top panel of Figure
\ref{fig:keelepw}. The white noise level (measured at high
frequencies) in this spectrum is only $\sim$230\,m\,s$^{-1}$. These
data have been weighted by the inverse square of their velocity error.
Once again the
dominant frequency is at 2075.8\,$\mu$Hz, with a velocity amplitude of
$\sim$13.5\,km\,s$^{-1}$. From our preliminary frequency analysis, we
have detected 17 frequencies with a S/N of 4 or better. Of these, one
frequency is 0.12\,$\mu$Hz away from the 2075.8\,$\mu$Hz peak. Since
this is less than the frequency resolution ($\sim$0.4\,$\mu$Hz), we
must question its reality. There are no noticeable problems
caused by the 2 week gap (which causes a splitting of
$\sim$0.8\,$\mu$Hz). Two of the frequencies detected are combination
frequencies, and if we relax our detection threshold of S/N=4 a
little, we find a further combination frequency. Just what causes
these combination frequencies is uncertain, although nonlinear effects
caused by the rapid rotation of PG\,1605+072 is one possibility. Four
of the frequencies we have measured have not been seen before in
velocity or photometry, so simulations will be required to determine
whether they are real. 

\section{Some Problems}
\label{sec:disc}

Some of the other problems encountered before, during and after the
campaign have already been mentioned (the small amount of 4\,m
spectroscopy, bad weather at La Silla and Siding Spring
Observatories), but the main problem has been the timing of
the photometric observations. These errors create a kind of paranoia
when it comes to dealing with low amplitude peaks very nearby (within
$\sim$1\,$\mu$Hz) high amplitude ones. Which peak is due to amplitude
variation, which is due to close mode spacing, and which is due to
timing problems? Detailed simulations will hopefully answer these
questions.

There are two possible ways to solve the timing problems, by
manual iteration or by examining phases. The former method involves
selecting sites with trustworthy times, systematically shifting the
times of one of the affected data sets until the combination of the
trustworthy data and the shifted data gives maximum amplitude. The
second method consists of determining the frequencies and phases of the
trustworthy data, fitting these frequencies to the affected data sets
(all with common time zeropoint), comparing the phases, and then
adjusting the times by the phase differences. This has been crudely
done already to determine which sites had timing problems in the first
place.

\section{Conclusions and Future Work}

There is still plenty of work to do before a proper and detailed
analysis of the MSST observations can be done. Reduction of the
photometry and analysis of the 4\,m spectroscopy needs to be
completed. The timing problems need to be investigated and then
corrected for, after which combination of all photometric (MSST+WET)
data can be done. Only then, frequencies, amplitudes and phases from
photometry and spectroscopy can be compared, and the identification of
modes in PG\,1605+072 can begin in earnest. We will call on the
pulsation theorists to help explain some of our results. 

We add a final comment on the feasibility of an MSST-like campaign on
other pulsating sdBs. We have shown the feasibility of this kind of
campaign beyond doubt for a bright sdB star with relatively long
periods, but what about other targets? There are several other
potential candidates for time-series spectroscopy, although they are
typically fainter and/or have shorter periods. These include
KPD\,2109+4401, Feige 48 and PG\,1219+534, which
have less complicated amplitude spectra, but are still bright enough
to observe, albeit with 4\,m telescopes only. These stars
have a lot fewer modes than PG\,1605+072, but this is actually
advantageous when looking for, and analysing, line profile
variations. So in the future look out for MSST II!

\section*{References}
Falter S., Heber U., Dreizler S., Schuh S.~L., Cordes O.-M. \&
Edelmann \indent H., 2003 A\&A, 401, 289

\noindent Heber et al., 2003, 13th European Workshop on White
Dwarfs. NATO-\indent ARW Workshop Series, p.\ 105

\noindent Kawaler S.~D., 1999 in: Proc.\ of the 11th European Workshop
on White \indent Dwarfs, eds Solheim J.-E. \& Meistas E.~G., ASP
Conf. Series vol. \indent 169, 158

\noindent Kilkenny D., Koen C., O'Donoghue D., et al., 1999 MNRAS, 303, 525

\noindent Koen C., O'Donoghue D., Kilkenny D., Lynas-Gray A.~E.,
Marang F. \indent \& van Wyk F., 1998 MNRAS, 296, 317

\noindent O'Toole S.~J., 2003, PhD thesis, University of Sydney

\noindent O'Toole S.~J., Bedding T.~R., Kjeldsen H., et al., 2000 ApJ,
537, L53

\noindent O'Toole S.~J., Bedding T.~R., Kjeldsen H., Dall T.~H. \& Stello D.,
2002 \indent MNRAS, 334, 471

\noindent O'Toole S.~J., J{\o}rgensen M.~A.~S.~G., Bedding T.~R., Kjeldsen H.,
Dall \indent T.~H., \& Heber U., 2003 MNRAS, 340, 856

\noindent Schuh S.~L., Dreizler S., Deetjen J.~L., G\"ohler, E., 2003
Baltic Astron., \indent 12, 167

\noindent Woolf V.~M., Jeffery C.~S. \& Pollacco D., 2002 MNRAS, 339, 497

\end{article}
\end{document}